\documentstyle[prl,aps,epsfig,amssymb,twocolumn]{revtex}

\begin{document}
\newcommand{\beq}{\begin{equation}}\newcommand{\eeq}{\end{equation}}
\newcommand{\barr}{\begin{eqnarray}}\newcommand{\earr}{\end{eqnarray}}

\newcommand{\andy}[1]{ }

\def\txt{\textstyle}

\def\ask{\marginpar{?? ask:  \hfill}}
\def\fin{\marginpar{fill in ... \hfill}}
\def\note{\marginpar{note \hfill}}
\def\check{\marginpar{check \hfill}}
\def\discuss{\marginpar{discuss \hfill}}
\def\hh{\widehat}
\def\wtilde{\widetilde}
\newcommand{\bm}[1]{\mbox{\boldmath $#1$}}
\newcommand{\bmsub}[1]{\mbox{\boldmath\scriptsize $#1$}}
\newcommand{\bmh}[1]{\mbox{\boldmath $\hat{#1}$}}

\draft

\title{ Stability and instability in parametric resonance and quantum Zeno
effect }
\author{ P. Facchi,$^{1}$ H. Nakazato,$^{2}$ S. Pascazio,$^{1}$ J.
Pe\v{r}ina,$^{3,4}$ and J. \v{R}eh\'{a}\v{c}ek$^{3}$}

\address{$^{1}$Dipartimento di Fisica, Universit\`a di Bari
     and Istituto Nazionale di Fisica Nucleare, Sezione di Bari,
 I-70126 Bari, Italy \\
$^{2}$Physics Department, Waseda University,  Tokyo 169-8555,
Japan
\\ $^{3}$Department of Optics, Palack\'{y} University, 17. listopadu
50, 772~00 Olomouc, Czech Republic\\ $^{4}$Joint Laboratory of
Optics of Palack\'{y} University and  Phys. Inst. Czech Acad.
Sci., 17. listopadu 50,\\ 772~00 Olomouc, Czech Republic}

\date{\today}

\maketitle

\begin{abstract}
A quantum mechanical version of a classical inverted pendulum is
analyzed. The stabilization of the classical motion is reflected
in the bounded evolution of the quantum mechanical operators in
the Heisenberg picture. Interesting links with the quantum Zeno
effect are discussed.
\end{abstract}


An inverted pendulum is an ordinary classical pendulum initially
prepared in the vertical upright position
\cite{seminal,Arnold,Fenn}. This is normally an unstable system,
but can be made stable by imposing a vertical oscillatory motion
to the pivot. In a few words, when the pivot is accelerated
upwards the motion is unstable, while when it is accelerated
downwards the motion can be stable: the periodic switch between
these two situations can be globally stable or unstable depending
on the values of some physical parameters. In particular, when
the frequency of the oscillation is higher than a certain
threshold, the system becomes stable. This result is a bit
surprising at first sight, but can be given an interesting
explanation in terms of the so-called parametric resonance
\cite{Arnold}.

In this Letter we shall study a system that can be viewed as a
quantum version of the inverted pendulum. The system to be
considered makes use of down-conversion processes interspersed
with zones where a linear coupling takes place between the
down-converted photon modes. It is similar to other examples
previously analyzed \cite{luisperina,RPFPM} in the context of the
quantum Zeno effect \cite{QZE}, where the ``measurement" is
performed by a mode of the field on another mode. When the
coupling between the two modes is large enough, the measurement
becomes more effective and the dynamics gets stable: this is just
a manifestation of the quantum Zeno effect, which consists in the
hindrance of the quantum evolution caused by measurements. The
very method of stabilization of the quantum system analyzed here
is one of its most interesting features and the configuration we
discuss is experimentally realizable in an optical laboratory. It
is therefore of interest both for the investigation of the
stable/unstable borderline for classical and quantum mechanical
systems and their links with the quantum Zeno effect.

We consider a laser field (pump) of frequency $\omega_p$, propagating
through a nonlinear coupler. The field is  considered to be classical
and the signal and idler modes are denoted by $a$ and $b$,
respectively. We will assume that all modes are monochromatic and the
amplitudes of the fields inside the coupler vary little during an
optical period (SVEA approximation). The effective (time-dependent)
Hamiltonian reads ($\hbar$=$1$)
\andy{hamilttot}
\begin{equation}\label{hamilttot}
H(t)=\omega_a a^{\dagger}a+ \omega_b b^{\dagger}b+ H_{\rm int}(t),
\end{equation}
where the interaction Hamiltonian is given by
\andy{hamiltint}
\begin{equation}\label{hamiltint}
H_{\rm int}(t) = \left\{
\begin{array}{ll}
\Gamma (a^{\dagger} b^{\dagger} e^{-i\omega_p t} + a b
e^{i\omega_p t}) & \mbox{if $0<t<\tau_1$}, \\ \Omega(a^{\dagger}b+
a b^\dagger) & \mbox{if $\tau_1<t<\tau_1+\tau_2$}
\end{array}
\right.
\end{equation}
and $H_{\rm int}(t+nT)=H_{\rm int}(t)$, with a period
$T=\tau_1+\tau_2$. The nonlinear coupling constant $\Gamma$ is
proportional to the second-order nonlinear susceptibility of the
medium $\chi(2)$ \cite{hong85}, $\Omega$ to the overlap between the
two modes \cite{coupl} and $n=0,1,\cdots,N$ is an integer.

\begin{figure}
\centerline{\epsfig{file=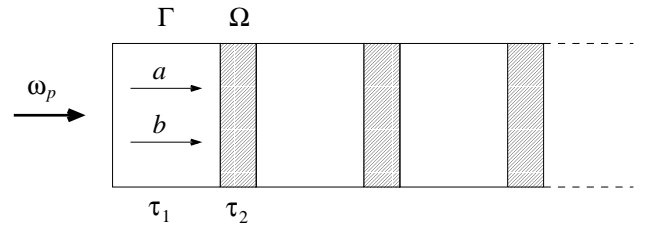,width=8.3cm}}
\caption{The system}
\label{fig:arnapp}
\end{figure}

We require the matching conditions $\omega_p=\omega_a+\omega_b$
and $\omega_a=\omega_b$ \cite{matching}. The above Hamiltonian
describes phase-matched down-conversion processes, for
$nT<t<nT+\tau_1$, interspersed with linear interactions between
signal and idler modes, for $n T+\tau_1<t<(n+1)T$. Since time is
equivalent, within our approximations, to  propagation length,
our system can be thought of as a nonlinear crystal cut into $N$
pieces, in each of which $a,b$ photons are created in a
down-conversion process. A similar configuration was considered in
\cite{luisperina}. Between these pieces, no new photons are
created by the laser beam, but the idler and signal modes
(linearly) interact with each other, for instance via evanescent
waves. See Fig.\ \ref{fig:arnapp}.
By introducing the slowly varying operators $a'=e^{i\omega_a t}a$,
$b'=e^{i\omega_b t}b$, the free part of the Hamiltonian
(\ref{hamilttot}) is transformed away and the Hamiltonian becomes
(suppressing all primes for simplicity)
\andy{hamiltintprime}
\begin{equation}\label{hamiltintprime}
H(t) = \left\{
\begin{array}{ll}
H_{\rm u} \equiv \Gamma( a^{\dagger} b^{\dagger} + a b) & \mbox{if
$0<t<\tau_1$}, \\ H_{\rm s} \equiv \Omega (a^{\dagger}b+a
b^\dagger) & \mbox{if $\tau_1<t<\tau_1+\tau_2$},
\end{array}
\right.
\end{equation}
with $H(t+nT)=H(t)$, yielding the equations of motion
\andy{eqmot}
\begin{equation}\label{eqmot}
\dot{a}=-i [a,H],  \qquad \dot{b}=-i [b,H].
\end{equation}
In terms of the variables
\andy{newvar}
\barr
x_\pm &=&\frac{1}{2}[(a+a^\dagger)\mp(b+b^\dagger)], \nonumber\\
p_\pm &=&-\frac{i}{2}[(a-a^\dagger)\mp(b-b^\dagger)],
\label{newvar}
\earr
which satisfy the equal-time commutation relations
$[x_+,p_+]=[x_-,p_-]=i$, others${}=0$,
the Hamiltonians become
\andy{newH12}
\barr
H_{\rm u} &=& \frac{\Gamma}{2} [(p_+^2-x_+^2) - (p_-^2-
x_-^2)],
\nonumber\\
H_{\rm s} &=& \frac{\Omega}{2} [(p_+^2+ x_+^2) -
(p_-^2+ x_-^2)].
\label{newH12}
\earr
They describe two uncoupled oscillators, whose equations of motion
are
\andy{equamotu,equamots}
\barr
& &
\left\{
\begin{array}{l}
\dot x_\pm = -i[x_\pm,H_{\rm u}]=\pm \Gamma p_\pm \\
\dot p_\pm = -i[p_\pm,H_{\rm u}]=\pm\Gamma x_\pm
\end{array}
\right.
\Longleftrightarrow \left\{
\begin{array}{l}
\ddot x_\pm -\Gamma^2 x_\pm=0 \\
\ddot p_\pm -\Gamma^2 p_\pm=0
\end{array}
\right. ,
\nonumber \\ 
& &  \nonumber \\
& &
\left\{
\begin{array}{l}
\dot x_\pm = -i[x_\pm,H_{\rm s}]=\pm\Omega p_\pm \\
\dot p_\pm = -i[p_\pm,H_{\rm s}]=\mp\Omega x_\pm
\end{array}
\right.
\Longleftrightarrow
\left\{
\begin{array}{l}
\ddot x_\pm +\Omega^2 x_\pm=0 \\
\ddot p_\pm +\Omega^2 p_\pm=0
\end{array}
\right. .
\label{equamots}
\earr
The first set of equations describes an {\em unstable} motion, the
second set a {\em stable} one, around the equilibrium point $x=p=0$.
Notice that the motion of $(x_-,p_-)$ is the time-reversed version of
that of $(x_+,p_+)$. This is due to the fact that the two
motions are governed by Hamiltonians with opposite sign in Eq.\
(\ref{newH12}). Henceforth, we shall concentrate on the variables
$(x_+,p_+)$ [the stability condition for $(x_-,p_-)$ is identical].
The solutions are
\andy{solunstable}
\barr
& & \pmatrix{x_+(\tau_1) \cr p_+(\tau_1)} = A_{\rm u} \pmatrix{x_+(0)
\cr p_+(0)},
\nonumber\\
& & \qquad \qquad A_{\rm u} \equiv \pmatrix{\cosh(\Gamma \tau_1)
& \sinh(\Gamma \tau_1) \cr
 \sinh(\Gamma \tau_1) & \cosh(\Gamma \tau_1)},
\label{solunstable}
\earr
for the period governed by $H_{\rm u}$ and
\andy{solstable}
\barr
& &\pmatrix{x_+(\tau_2) \cr p_+(\tau_2)} = A_{\rm s} \pmatrix{x_+(0)
\cr p_+(0)},
\nonumber\\
& & \qquad \qquad
A_{\rm s} \equiv \pmatrix{\cos(\Omega \tau_2) &
\sin(\Omega \tau_2) \cr -\sin(\Omega \tau_2) & \cos(\Omega
\tau_2)}, \label{solstable}
\earr
for that governed by $H_{\rm s}$. Remember that $T=\tau_1+\tau_2$
is the period of the  Hamiltonian $H(t)$ in
(\ref{hamiltintprime}).

The dynamics engendered by (\ref{hamiltintprime}) at time $t=NT$
(remember that $n=1,\ldots,N$) yields therefore \andy{globdyn}
\barr
\pmatrix{x_+(NT) \cr p_+(NT)}
 = A^N \pmatrix{x_+(0) \cr p_+(0)}, \qquad A \equiv A_{\rm s} A_{\rm u} .
\label{globdyn}
\earr
These equations of motion have the same structure of a classical
inverted pendulum with a vertically oscillating point of suspension
\cite{Arnold}, whose classical map is given by the product of
two matrices $A_{\rm cl}\equiv A_2A_1$, with
\andy{A1A2}
\barr
A_1&\equiv&\pmatrix{\cosh(k_1\tau)
                            & k_1^{-1}\sinh(k_1\tau) \cr
                            k_1\sinh(k_1\tau)
                            & \cosh(k_1\tau)},
\nonumber\\
A_2&\equiv&\pmatrix{\cos(k_2\tau)
                            &k_2^{-1}\sin(k_2\tau) \cr
                            -k_2\sin(k_2\tau)
                            & \cos(k_2\tau)},
\label{A1A2}
\earr
where the parameters $k_1$ and $k_2$ are subject to the physical
condition $k_1>k_2>0$.  Observe that our system has more freedoms:
$\tau_1$ and $\tau_2$ are in general different and the parameters
$\Omega$ and $\Gamma$ do not have to obey any additional constraint.

\begin{figure}
\centerline{\epsfig{file=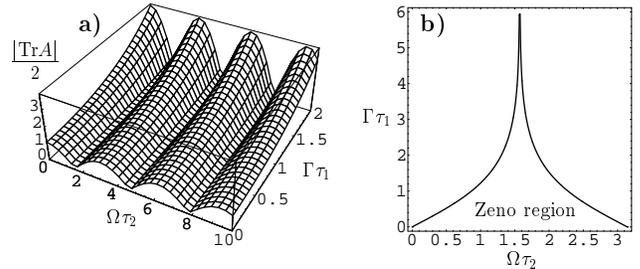,width=8.3cm}}
\caption{Stability
condition (\ref{trac2}) in parameter space. a) $|\mbox{Tr}A|/2$ vs
$\Omega \tau_2$ and $\Gamma\tau_1$; b) Stability (Zeno) region.}
\label{fig:smallt}
\end{figure}

The global motion is stable or unstable, according to the value of
$|\mbox{Tr}A| \lesseqgtr 2$ \cite{Arnold}. The stability condition
$|\mbox{Tr}A|<2$ reads
\andy{trac2}
\beq
|\mbox{Tr}A|/2=|\cos(\Omega \tau_2) \cosh(\Gamma \tau_1)|<1.
\label{trac2}
\eeq
This relation is of general validity and holds for any value of the
parameters $\Omega, \Gamma$ and $\tau_i$. The value of
$|\mbox{Tr}A|/2$ is shown in Fig.\
\ref{fig:smallt}a). A small-$\tau$ expansion (the physically relevant regime:
see final discussion) yields
\andy{tract}
\beq
1 - (\Omega^2\tau_2^2-\Gamma^2\tau_1^2)/2 +
O(\tau^4) <1, \label{tract}
\eeq
so that the system is stable for $\Omega \tau_2>\Gamma \tau_1$
when $\tau_2\to0$.

It is interesting to discuss the stability condition just obtained
for the $(x,p)$ variables in terms of the number of down-converted
photons. To this end, let us look at some limiting cases.
[Needless to say, the analysis could be done from the outset in
terms of $n_a$ and $n_b$ and would yield an identical stability
condition (\ref{trac2}).] When $\Omega=0$ in
(\ref{hamiltintprime}) and following equations, only the
down-conversion process takes place and both $n_a=a^\dagger a$
and $n_b=b^\dagger b$ grow exponentially with time. There is an
exponential energy transfer from the pump to the $a,b$ modes. On
the other hand, if $\Gamma=0$ and the system is prepared in any
initial state (except vacuum, whose evolution is trivial), $n_a$
and $n_b$ oscillate in such a way that their sum is conserved
(this is due to the property $[n_a+n_b,H_{\rm s}]=0$). If both
$\Omega$ and $\Gamma$ are nonvanishing, these two opposite
tendencies (exponential photon production and bounded
oscillations) compete in an interesting way. When
$\Gamma\tau_1>\Omega\tau_2$, in the limit $\tau_1\to 0$, the
exponential photon production dominates and there is no way of
halting (or even hindering) this process: the (external) pump
transmits energy to the $a,b$ modes. In terms of the $(x,p)$
variables, the stability condition (\ref{tract}) {\em cannot} be
fulfilled and the oscillator variables move exponentially away
from the origin. The opposite situation
$\Omega\tau_2>\Gamma\tau_1$ is very interesting and displays some
quite nontrivial aspects: The motion becomes stable and the pump
does not transmit energy to the $a,b$ modes anymore (the two
modes oscillate).

In general and for arbitrary values of all parameters, the action
of $H_{\rm s}$ can be viewed as a sort of measurement
\cite{RPFPM,MPS,comment}, in the following sense: the $a$ mode
performs an observation on the $b$ mode and {\it vice versa}, the
photonic states get entangled and information on one mode is
encoded in the state of the other one. For example, the condition
$\Omega\tau_2=\pi/2$ yields an ``ideal measurement" of one mode
on the other one, for in such a case the states $|1_a,0_b\rangle
\leftrightarrow |0_a,1_b\rangle$ evolve into each other. From
this viewpoint, the stabilization regime just investigated can be
considered as a quantum Zeno effect \cite{QZE}, in that the
measurements essentially affect and change the original dynamics.
In fact, if one considers $\Omega\tau_2$ as the ``strength" of
the measurement, by increasing (at fixed $\Omega\tau_2$) the
frequency of measurements, i.e. by letting $\tau_1\to0$, the
system moves down along a vertical line in Fig.\
\ref{fig:smallt}b) and enters a region of stability (Zeno region)
from a region of instability. (Notice that it is not necessary to
consider the $\tau_1=0$ limit (``continuous measurement") in
order to stabilize the dynamics; there is a threshold, given by
the curve in Fig. 2 b), at which stability and instability
interchange.) Analogously, at fixed $\Gamma\tau_1$, by moving
along a horizontal line $\Omega\tau_2\to \pi/2$ the system enters
a region of stability because the measurement becomes more
``effective:" indeed, as emphasized before, $\Omega\tau_2=\pi/2$
is a $\pi$-pulse condition and leads to a very effective
measurement of one mode on the other one. It is worth stressing
that even an instantaneous measurement (projection) can be
obtained by letting $\tau_2\to0$, while keeping $\Omega \tau_2$
finite (the so-called impulse approximation in quantum mechanics),
and in this case our system yields the standard formulation of
the quantum Zeno effect.

It is interesting (and convenient from an experimental
perspective) to consider a single-mode version of the Hamiltonian
(\ref{hamiltintprime}), in which the down-conversion process is
replaced by a sub-harmonic generation process (degenerated
parametric down conversion). The single-mode effective Hamiltonian
reads
\andy{hamilt}
\begin{equation} \label{hamilt}
H(t)=\omega a^{\dag}a + H_{\rm int}(t),
\end{equation}
where the interaction Hamiltonians describing the unstable and stable
part of the device are
\andy{int}
\begin{equation} \label{int}
H_{\rm int}=\left\{
\begin{array}{ll}
(\Gamma/2)(a^{\dag 2}e^{-2i\omega t}+a^2e^{2i\omega t}) & \mbox{if $0<t<\tau_1$},\\
(\Omega/2)(a^{\dag}a+a a^\dagger) & \mbox{if
$\tau_1<t<\tau_1+\tau_2$},
\end{array}\right.
\end{equation}
respectively and $H_{\rm int}(t+nT)=H_{\rm int}(t)$. By
introducing the slowly varying operator $a'=e^{i\omega_a t}a$,
the free part of the Hamiltonian (\ref{hamilt}) is transformed
away and the Hamiltonian becomes (suppressing again all primes)
\andy{hamiltintprime1}
\begin{equation}\label{hamiltintprime1}
H(t) = \left\{
\begin{array}{ll}
H_{\rm u} \equiv (\Gamma/2) ( a^{\dagger 2}+ a^2 ) & \mbox{if
$0<t<\tau_1$}, \\ H_{\rm s} \equiv (\Omega/2) (a^{\dagger}a+a
a^{\dagger}) &  \mbox{if $\tau_1<t<\tau_1+\tau_2$},
\end{array}
\right.
\end{equation}
under which the equation of motion $\dot{a}=-i [a,H]$ follows.

In terms of the variables $x=(a+a^{\dag})/\sqrt{2},
p=-i(a-a^{\dag})/\sqrt{2}$ the Hamiltonians read
\andy{hamilttrans}
\beq
\label{hamilttrans}
H_{\rm u}=\frac{\Gamma}{2}(x^2-p^2), \quad  H_{\rm s}=\frac{\Omega}{2}(x^2+p^2).
\eeq
These Hamiltonians are identical to the two-mode versions
(\ref{newH12}) describing the decoupled mode $(x_+,p_+)$, apart
from the substitution $\Gamma\to-\Gamma$. Hence, the stability
condition is given again by Eq.\ (\ref{trac2}), which is even in
$\Gamma$. Also in this case one can talk of quantum Zeno, but the
``measurement" is performed by the single mode on itself.

\begin{figure}
\centerline{\epsfig{file=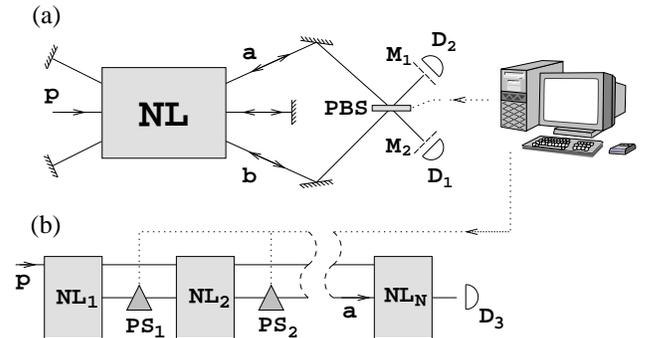,width=8.3cm}}
\caption{Experimental setup. (a) Possible experimental realization
of the Hamiltonian (\ref{hamilttot})-(\ref{hamiltint}). NL,
nonlinear crystal; M$_{i}, (i=1,2)$, semitransparent mirrors;
D$_{i}$, detectors; PBS, polarizing beamsplitter. (b) Possible
experimental realization of the Hamiltonian
(\ref{hamilt})-(\ref{int}). NL$_{i} (i=1,\ldots,N)$, nonlinear
crystals; PS$_{i} $, phase shifters; D$_3$, detector. The dotted
lines indicate which elements are computer controllable.}
\label{fig:exptsetup}
\end{figure}

It is interesting to discuss a possible experimental realization
of the two situations considered in this Letter. The experimental
arrangement sketched in Fig.\ \ref{fig:exptsetup}(a) corresponds
to the two-mode (nondegenerate) case, whereas that sketched in
Fig.\ \ref{fig:exptsetup}(b) to the single-mode (degenerate)
case. In Fig.\ \ref{fig:exptsetup}(a) a type II down-conversion
process generates two orthogonally polarized beams of
down-converted light of the same frequency. The two beams are
mixed using a polarizing beamsplitter PBS. The stable part of the
evolution of the system is realized by two successive passes of
the beams through the beamsplitter. Its reflection coefficient,
and hence $\Omega\tau_2$, is adjusted by rotating it. Mirrors and
semitransparent mirrors keep sending the beams through the crystal
many times. A successful stabilization of the unstable system is
manifested in the decrease of the rate of photon registrations at
detectors $D_1$, $D_2$ at a certain position of the beamsplitter
PBS. A different setup is sketched in Fig.\
\ref{fig:exptsetup}(b), where $N$ processes of subharmonic
generation take place in $N$ nonlinear crystals with controlled
phase shifters in between them. For appropriately chosen phase
shifts $\theta_i=(\Omega\tau_2+C_i) \bmod 2\pi$, where $C_i$ are
$N-1$ phase shifts intrinsic to the actual experimental
arrangement (given by distances between crystals, etc.), the
generation of the subharmonic wave is suppressed.

In order to give a reasonable estimate of the value of the
coupling constant $\Gamma$, consider that, due to the
correspondence principle, the gain of classical and quantum
parametric amplifiers must be the same; therefore one can use the
well-known classical formula for the nonlinear coupling parameter
$\Gamma_c$ governing the space evolution inside the nonlinear
medium, which in MKS units reads
\andy{gamma}
\begin{equation} \label{gamma}
\Gamma_c^2=\frac{\eta^3}{2}\chi(2)^2\omega_a\omega_b I_p.
\end{equation}
Here $\eta$ is the impedance of the medium, $\chi(2)$ is the
second-order susceptibility, $\omega_a$ and $\omega_b$ are the
frequencies of modes $a$ and $b$, respectively, and $I_p$ is the
intensity of the pump beam. The following numerical values could
be typical for a performed experiment: $\eta \approx 220\Omega$,
$\chi(2)\approx 2\times10^{-23}$ CV$^{-2}$, $\omega_a=\omega_b
\approx3\times10^{15}$s$^{-1}$ and $I_p\approx 10^5$Wm$^{-2}$.
Hence the nonlinear coupling parameter is of the order of
$\Gamma_c\approx 0.1$m$^{-1}$. Reasonable lengths of nonlinear
crystals are of the order of $l\approx 10^{-2}$m, so that the
dimensionless product of interest can be estimated to be about
\andy{product}
\begin{equation} \label{product}
\Gamma\tau_1=\Gamma_c l \approx 0.001.
\end{equation}
This means that the down-converted beam(s) ought to pass the
nonlinear region many times in order to show an explosive
increase of its (their) intensity(ies). This could be achieved by
placing the nonlinear crystal in a resonator as shown in Fig.\
\ref{fig:exptsetup}(a). However, in order to observe a
significant change of the dynamics of the process in question due
to the performed stabilization, a few passes might already turn
out to be sufficient.

In conclusion, we have discussed a striking quantum-optical
analogue of a well-known classical unstable system. By
interspersing the nonlinear regions with regions of suitably
chosen linear evolution, the global dynamics of our system can
become stable and the generation of down-converted light can be
strongly suppressed. This behavior has an interesting
interpretation in terms of the quantum Zeno effect: by increasing
the ``strength" of the observation performed by the $a$ mode on
the $b$ mode and {\it vice versa}, in the sense discussed before,
the evolution is frozen and the system tends to remain in its
initial state. This phenomenon is somewhat counterintuitive: in
the setups in Fig.\ \ref{fig:exptsetup}, even though the beams
are forced to go through the crystal many times, no exponential
photon production takes place. The experiment seems feasible and
its realization would illustrate an interesting aspect related to
the stabilization of a seemingly explosive behavior.

\acknowledgments

We thank Ond\v{r}ej Haderka, Martin Hendrych and Zden\v{e}k
Hradil for helpful discussions. We acknowledge support by the
TMR-Network ERB-FMRX-CT96-0057 of the European Union, by Grant
VS96028 and Research Project CEZ:J14/98 ``Wave and particle
optics'' of the Czech Ministry of Education (J.P.\ and J.\v{R}.),
by the internal grant by Palack\'{y} University (J.\v{R}.) and by
a Grant-in-Aid for Scientific Research (B) (No.\ 10044096) from
the Japanese Ministry of Education, Science and Culture (H.N.).


\begin{thebibliography}{99}
\bibitem{matching} \andy{matching}
The latter matching condition can be relaxed at the cost of
introducing a second classical pump wave of frequency
$\omega_p'=\omega_a-\omega_b$ and the Hamiltonian $\Omega
(a^{\dag}be^{-i\omega_p't}+ ab^{\dag}e^{i\omega_p't})$, for
$\tau_1<t<\tau_1+\tau_2$, in (\ref{hamiltint}). Physically, this
corresponds to replacing the linear exchange between modes $a$ and
$b$ with the nonlinear process of difference frequency generation.
\end{thebibliography}
\end{document}